\documentstyle[aps,prc,twocolumn,epsfig]{revtex}

\begin{document}
\wideabs{\ 
\author{P.Balenzuela, A. Chernomoretz and C.O.Dorso}
\title{Time dependence of critical behavior in multifragmentation}
\address{Departamento de F\'{\i}sica, Facultad de Ciencias Exactas y\\
Naturales Universidad de Buenos Aires\\
Pabellon I, Ciudad Universitaria, Nu\~nez\\
1428 Buenos Aires, Argentina}
\date{\today}
\maketitle

\begin{abstract}
We study signatures of critical behavior in microscopic simulations of
small, highly excited Lennard-Jones drops. We focus our attention on the behavior of 
the system at the time of fragment formation (which takes place in phase space)
and compare the results with the corresponding ones obtained at asymptotic times
(experimentally accessible). The four critical exponents ($\tau$,$\beta$, 
$\sigma$ and $\gamma$) found at fragmentation time have shown to be stable
against time evolution, indicating that the asymptotic stage reflects
accurately the physics at fragmentation time. Even though 
evidence of critical behavior arises from the calculations, we can not
affirm that the system is performing a second order like phase transition.
\end{abstract}

}
\bigskip

\newpage The process of rapid expansion and fragmentation of highly excited
drops attracts the attention of physicists in different areas, in
particular in the field of nuclear physics. The possibility of facing a
phase transition was initially triggered by the work of the Purdue Group 
\cite{purdue} when the resulting fragment mass spectra of proton-nucleus
colission was fitted by a power law like dependence. This has initiated a series of 
works focusing on the extraction of signals of critical behavior from both,
experimental measurements as well as data resulting from numerical
simulations \cite{gilkes,latobona,nuovochi,dorlo,philip,elliot2,nato}. In
the experimental case the only information available corresponds to the
asymptotic regime and the primordial fragments have to be reconstructed from
this data by the application of different approximations \cite{dagostino}.
In a series of works (see for example \cite{elliot}, \cite{bauer}) critical
exponents have been calculated from the analysis of the experimental
fragment mass distributions. On the other hand, in the case of numerical
simulations all the microscopic information is available at all times, and 
if  appropriate fragment recognition algorithms are used, one can unveil
the complex fragmentation mechanism. In this work we use the already
presented Early Cluster Recognition Algorithm (ECRA) to recognize fragments 
\cite{ecra93}. The advantage of this methodology is that it is able to find, very
early in the evolution, the partitions of the system that give rise to the 
asymptotic fragments. These partitions (ECRA clusters) are the most bound
density fluctuations in phase space and has been proved to be the seeds of
the asymptotic fragments \cite{strador,strador2}. As all order correlations are 
available, it is possible to follow the evolution in time of the ECRA clusters 
and determine the time of fragment formation ($\tau_{ff})$ as the time at which 
these clusters attain microscopic stability. Once the distribution of ECRA clusters 
at $\tau_{ff}$ is known, we can look for signals of critical behavior and extract the
values of the critical exponents $\tau$, $\beta$, $\sigma$ and $\gamma$ (for
a definition of these critical exponents see section III) . We then test the
stability of the values of these exponents as a function of time, checking
if the values extracted at asymptotic times are similar to the ones obtained
at $\tau_{ff}$. In this way, we try to determine if the asymptotic stage 
information properly characterize the state of the system at fragmentation time.

This work is organized as follows: In Section I we describe the model used
to simulate the fragmentation process as well as the algorithms adopted to 
recognize the fragments. In Section II we will give a brief
description of the scaling model applied to analyze the data extracted from
simulations, as well as the behavior of the system near the critical point. In
Section III we analyze the behavior of some signals of critical behavior
at $\tau _{ff}$ and at asymptotic time. The selected signals of critical
behavior are: the second moment of the fragment distribution ($M_{2}$), the
normalized variance of the mass of the maximum fragment ($NVM$), the best
power law fit to the fragment spectra according to the assumed scaling model 
(minimum $\chi ^{2}$) and the critical exponents $\tau$, $\gamma$, $\beta$ and 
$\sigma$. Finally, in Section IV, conclusions are drawn.

\section{Computer Experiments}

In previous works \cite{strador,strador2,chernodor} we have already analyzed
the dynamics of fragment formation and the caloric curve for excited drops
made up of 147 particles interacting via a 6-12 Lennard Jones potential. The
interaction potential reads:

\begin{equation}
V(r)=\left\{ 
\begin{array}{ll}
4\epsilon\left[\left(\frac{\sigma }{r}\right)^{12}-\left(\frac{\sigma}{r}%
\right)^{6}-\left(\frac{\sigma }{r_{c}}\right)^{12}+\left(\frac{\sigma}{r_{c}%
}\right)^{6}\right] & r<r_c \\ 
0 & r \ge r_c .
\end{array}
\right.
\end{equation}

\smallskip We took the cut-off radius as $r_{c}=3\sigma $. Energy and
distance are measured in units of the potential well ($\epsilon $) and the
distance at which the potential changes sign ($\sigma $), respectively. The
unit of time used is: $t_{0}=\sqrt{\sigma ^{2}m/48\epsilon }$. In our
numerical experiments initial conditions where constructed using the already
presented \cite{strador,strador2} method of cutting spherical drops composed
of 147 particles out of equilibrated, periodic, 512 particles per cell L.J.
system. 
One of the key ingredients in the analysis of fragmentation is the
determination of the time at which fragments are formed. This can be
accomplished if a proper fragment recognition method is employed. In this
paper, clusters are defined according to the ECRA algorithm \cite{ecra93} in
which the fragments are related to the most bound density fluctuations in
phase space, as was mentioned above. These fragments are referred as ECRA
clusters. In this case fragments are given by the set of clusters $\{C_{i}\}$
for which the sum of the fragment internal energies attains its minimum
value:

\begin{eqnarray}
{\{C_{i}\}} &{=}&{\hbox{min}}_{{\scriptstyle\{C_{i}\}}}{\textstyle {%
[E_{\{C_{i}\}}=\sum_{i}E_{int}^{C_{i}}]}}  \nonumber \\
E_{int}^{C_{i}} &=&[\sum_{j\in C_{i}}K_{j}^{c.m.}+\sum_{{j,k\in C_{i}%
}{j\le k}}V_{j,k}]  \label{eq:eECRA}
\end{eqnarray}

where the sum in the first equation of (\ref{eq:eECRA}) is over the clusters of the
partition, $K_{j}^{c.m.}$ is the kinetic energy of particle $j$ measured in
the center of mass frame of the cluster which contains particle $j$, and $%
V_{ij}$ stands for the inter-particle potential.

Another way of recognizing fragments is provided by the Minimum Spanning
Tree (MST) algorithm\cite{strador,strador2}. In this case, given two
particles $i,j,$ they belong to a cluster $C$ if the following relation is
satisfied:

\begin{equation}
i\in C\Longleftrightarrow \exists j\in C\text{ }/\text{ }r_{ij}\text{ }\leq
r_{cl}
\end{equation}

with $r_{ij}$ the interparticle distance and $r_{cl}$ the clusterization
radius ( $r_{cl}\leq r_{c}=3\sigma )$. In our case, we took $r_{cl}=r_{c}$.
It should be noticed that in this definition the correlations in momentum
space between particles are completely disregarded.

The asymptotic fragments detected by MST algorithm are observables. On the
other hand, the ECRA clusters, being defined in phase space, are observables
only when they coincide with MST clusters. For earlier times they are
usually embedded into the MST clusters. It has been shown \cite
{ecra93,strador,strador2,aichelin} that different properties of the ECRA
fragments (for instance the average size of the maximum fragment, or the
average multiplicity of fragments in given bins, etc.) become stable very
early in the evolution, much earlier than the times of stabilization of the
same quantities calculated from MST clusters. As already mentioned in the
introduction, the time of fragment formation is associated to the time at
which the ECRA clusters attain microscopic stability. This occurs when the
system switches from a regime dominated by fragmentation to one in which the
dominant decay mode is evaporation. In order to estimate this time 
($\tau_{ff}$) we use the so called, Short Time Persistence (STP), in which we
calculate the stability of the ECRA clusters against
evaporation-fragmentation and coalescence. It is defined in the following
way: at a given time $t$ we analyze each fragment $C_{i}^{t}$ of size $%
N_{i}^{t}$ by searching on all the fragments $C_{j}^{t+dt}$ present at time $%
t+dt$ for the biggest subset $N_{\max }]_{i}^{t+dt}$ of particles that
belonged to $C_{i}^{t}.$ Then we assign to this fragment a value $STP_{d}=%
\frac{N_{\max }]_{i}^{t+dt}}{N_{i}^{t}}$. This term account for the
''evaporation-fragmentation process'', but on the other hand one has to take
care of the cases in which the $N_{\max }]_{i}^{t+dt}$ does not constitute a
free cluster but is embedded in a bigger fragment of mass $N_{i}^{t+dt}$. We
include this effect by defining $STP_{i}=\frac{N_{\max }]_{i}^{t+dt}}{%
N_{i}^{t+dt}}$. Finally the Short Time Persistence reads :

\smallskip 
\begin{equation}
STP(t,dt)=\left\langle \left\langle \frac{STP_{d}(t,dt)+STP_{i}(t,dt)}{2}
\right\rangle _{m}\right\rangle _{e}
\end{equation}

where $\left\langle ...\right\rangle _{m}$ is the mass weighted average over
all the fragments with size $N>3$. And $\left\langle ...\right\rangle_{e}$
is the average over an ensemble of fragmentation events at a given energy $E$
or multiplicity $m$. From this definition it is clear that $STP \sim 1$ if
the analyzed partition remains unaltered during $dt$ and, on the other hand, 
$STP \sim 0$ if its microscopic composition presents large changes during
that lapse of time.

\smallskip In Fig. 1 we show one typical calculation of this magnitude.
This is for a very energetic case $E=2.65\epsilon $ for which a fast
exponentially decaying spectrum is obtained. In this figure we show 
$STP(t,dt)$ as a function of $t$ and for different values of $dt$, namely 
$dt=0.5t_{0},1t_{0}.,1.5t_{0},2t_{0}$. The almost horizontal line denotes the
reference value of $STP$ and corresponds to fragments undergoing only a
simple evaporation process. In this way we say that when $STP$ crosses this
line the behavior of the system goes, on the average, from fragmentation to
evaporation, and we call this time $\tau _{ff}$. Using the above mentioned
criteria we have obtained time-scales ranging from $\tau _{ff}\sim 75t_{0}$
for $E=-0.5\epsilon $ up to $\tau _{ff}\sim 10t_{0}$ for $E=3.7\epsilon $.

\begin{figure}
\centerline{\epsfig{figure=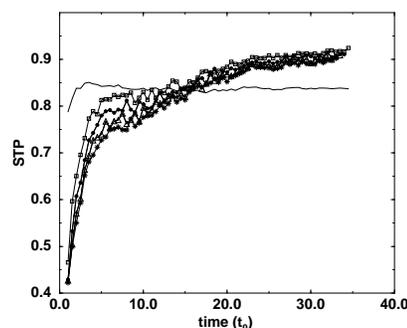,height=6cm,angle=-90}}
\caption{Short time persistence (STP) as function of time for four different
values of dt (empty squares:$dt=0.5t_{0}$, full circles $1t_{0}$, empty
triangles $1.5t_{0}$ and stars $2t_{0}$) and the reference value (solid
line)}.
\end{figure}

In the same way, but working with the fragments given by the MST algorithm we
get the times of fragment emission $\tau_{fe}$. ,i.e. the times at which the
clusters defined in q-space become stable. These times are much larger than
the corresponding time of fragment formation ($\tau_{ff}$). For example, for 
$E=-0.5\epsilon$, $\tau_{fe}=100t_0$. We took as asymptotic time, ($\tau_a$), 
a value much larger that all the $\tau_{fe}$ calculated, i.e., $\tau_a=600t_0$.

Once we determine the time of fragment formation ($\tau_{ff}$) and the asymptotic 
time ($\tau_{a}$), it is possible to calculate the corresponding 
ECRA fragment mass distribution at $\tau_{ff}$ and compare it with the 
corresponding MST fragment mass spectra at $\tau_a$. The result 
of such a calculation are displayed in Fig. 2. Three typical mass spectra are 
shown: U-shaped in panel (a), power law like in (b) and exponential decaying one 
in (c). 
Filled circles denote the ECRA distributions at time of fragment formation and
empty triangles the MST ones at asymptotic times. It can be
notice that the resulting distributions are esentially the same.

\begin{figure}
\centerline{\epsfig{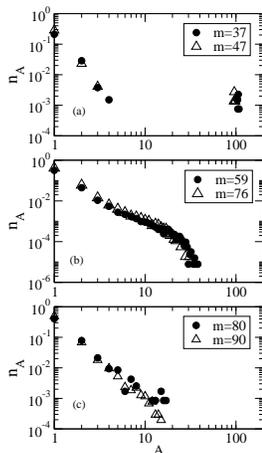}}
\caption{Three typical fragment distributions ($n_A$: number of clusters per particle 
of size $A$) corresponding to three different 
excitation energies and classified by their multiplicity values. 
Black circles correspond to ECRA fragment distributions at $\tau_{ff}$ and empty 
triangles to MST ones at asymptotic times. The values of $m$ indicate the 
multiplicity of the correspondig mass spectra . In panel (a) we show the U-shaped 
distributions (low excitation energies), in (b) the power law and in (c) the 
exponential decaying ones (high excitation energies)}.
\end{figure}

\section{The Scaling Model: critical exponents and other signatures of
criticality}

The behavior of the system near its critical point can be described in terms
of scaling laws. A widely used scaling model for the fragment mass
distribution \cite{elliot,bauer,stauffer}, based in renormalization group
arguments \cite{gulmi}, can be applied to this kind of problem.  In this
approach, the cluster distribution can be written as:

\begin{equation}
n_A(\epsilon)=q_0A^{-\tau}f(z),  \label{eq:fdm}
\end{equation}

where $n_{A}$ is the probability per particle of having a cluster of size $A$%
, $\tau$ is a critical exponent, $q_{0}$ a normalization constant and $f(z)$
a scaling function. This function depends on the distance to the critical
point, $\epsilon$, and the mass number, $A$, via the combination \cite
{stauffer}:

\begin{equation}
z=A^{\sigma}\epsilon ,  \label{eq:sigma}
\end{equation}

with $\sigma $ a critical exponent. The distance to the critical point can be 
defined as $\epsilon =(T_{c}-T)/T_{c}$ for usual thermodynamic systems, and
$\epsilon =(m_{c}-m)/m_{c}$ \cite{elliot} for multifragmentation experiments. 

At the critical point, $\epsilon=0$ (with $f(0)=1$ \cite{stauffer}), and the
cluster mass distribution becames a power law:

\begin{equation}
n_A=q_0A^{-\tau},  \label{eq:pl}
\end{equation}

where the normalization $q_0$ depends on $\tau$ via a Riemann $\zeta$
function \cite{stanley}:

\begin{equation}
q_0=\frac{1}{\sum_A A^{1-\tau}} .  \label{eq:q0}
\end{equation}

This dependence, coming from the normalization condition $M_{1}(\epsilon=0)=1$, 
should be taken into account to properly extract the critical exponent $\tau$ when
fitting the cluster mass distribution to a power law.

A relevant quantity in the analysis of critical behavior is the second
moment of the mass distribution:

\begin{equation}
M_{2}=\sum n_{A}A^{2}.  \label{eq:M2}
\end{equation}

It has been shown (see for example \cite{stauffer}) that $M_2$ diverges at
the critical point if the fragment distribution satisfies equation (\ref
{eq:fdm}). This divergence can be described in terms of the critical
exponent $\gamma$:

\begin{equation}
M_{2}(\epsilon )\propto \Gamma _{\pm }|\epsilon |^{-\gamma }.
\label{eq:gamma}
\end{equation}

Another relevant magnitude in the search of critical behavior is the
normalized variance of the size of the biggest cluster $(NVM)$ \cite{dorbona}%
, defined by:

\begin{equation}
NVM=\frac{<A_{max} - <A_{max}>>^2}{<A_{max}>}
\end{equation}

where $A_{max}$ is the size of the biggest fragment and $<...>$ is the
average over an ensemble of fragmentation events with a given multiplicity $%
m $. In \cite{dorbona} it was shown that it is a robust signature of
critical fluctuations.

It is worth noting that for finite systems it is expected that scaling assumptions 
would be valid in a range of masses where finite size effects can be avoid and 
where the behavior of the system resembles the behavior of the infinite one.

We will see in the next section that for exploding LJ drops, the fragment
distributions at $\tau_{ff}$ and $\tau_a$ follows the scaling model within a 
rather broad range of mass. 

\section{Criticality Analysis}

In what follows we search for critical behavior in our numerical simulations
via the analysis of the fragment mass disttributions at both, fragmentation time 
$(\tau_{ff})$ and asymptotic times $(\tau_{a}).$

\subsection{Signals of critical behavior}

We first calculate the normalized variance of the mass of the biggest
fragment, $NVM$, and the second moment of the cluster mass distributions, 
$M_{2}$ as a function of the multiplicity, being the multiplicity a relevant 
observable in multifragmentation. The biggest fragment of each event should be 
removed for the liquid phase ($m<m_c$) in the calculation of $M_2$ 
\cite{stauffer}. Because we don't know a priory the value of the critical 
multiplicity, the biggest fragment was excluded for all events \cite{elliot}.

The dependence of these magnitudes with the multiplicity is shown in Fig. 3
for the two times analyzed, $\tau _{ff}$ and $\tau _{a}$. It can be seen
that both observables peak at, essentially, the same multiplicity value for 
each of the analized times. At $\tau_{ff}$, the NVM peaks for $m=60\pm 1$
[Fig. 3(a)] and the $M_{2}$ for $m=65\pm 4$ [Fig. 3(b)] (black triangles in both
cases), whereas at asymptotic times, the magnitudes peak for $m=73\pm 1$
[Fig. 3(a)] and $m=77\pm 3$ [Fig. 3(b)] respectively (empty symbols). The higher
multiplicity values at asymptotic times are due to evaporation processes
undergone by the fragments, which increase the number of monomers and
dimers. If we consider only fragments larger than $A=2$, both magnitudes 
($M_2$ and $NVM$) peaks at the same value of multiplicity at both times.
These results are compatible with the presence of critical behavior. 

\vspace{0.8cm}
\begin{figure}
\centerline{\epsfig{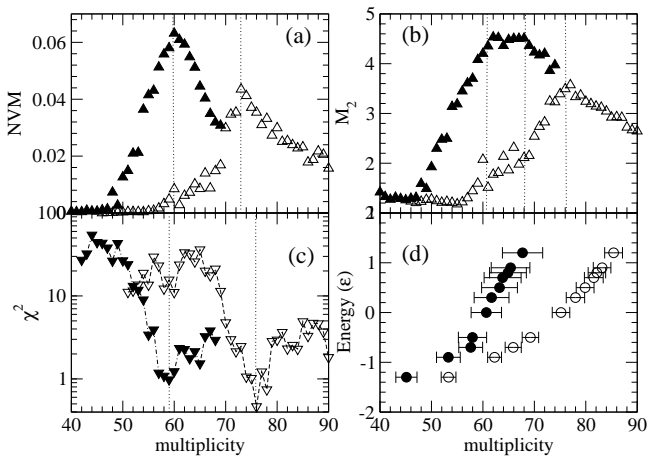}}
\caption{Normalized variance of the size of the maximum fragment (NVM) (a),
second moment of the cluster distribution ($M_2$) (b), and best fit to a 
power law (with $q_0=q_0(\tau)$ via eq.(\ref{eq:q0})) (c) as function of the
multiplicity for fragmentation time ($\tau_{ff}$), full triangles, and for
asymptotic time ($\tau_{a}$), empty triangles. In (d) we can see average
multiplicity and its standard desviation for each energy for $\tau_{ff}$
(full circles) and for $\tau_{a}$ (empty circles)}.
\end{figure}

We will now search for a power law mass spectra according to the scaling model 
depicted in equation (\ref{eq:fdm}).
As we have seen, the normalization $M_{1}(\epsilon=0)=1$ imposes a condition
on $q_{0}$ such that $q_{0}=q_{0}(\tau)$ via a Riemann $\zeta$ function 
(eq.\ref{eq:q0}). Therefore, if we want to find a power law behavior in the mass
spectra obtained from the numerical simulations, we must fit them with
equation (\ref{eq:pl}) taking into account that the normalization, $q_0$, is given
by equation (\ref{eq:q0}). This method was successfully used in \cite{elliot}
both, for percolation and $Au+C$ multifragmentation. In our case, the best fitted 
mass spectra by $f(A,\tau )=q_{0}(\tau )A^{-\tau }$ determines the
critical multiplicity whereas the corresponding slope in a log-log plot, the
critical exponent $\tau$. The quality of the fitting procedure was measured
via the standard $\chi^{2}$ coefficient (i.e., a minimum in $\chi^{2}$ 
corresponds to the best fit).

The fitting was performed in the mass range $0.02A_{tot}<A<0.15A_{tot}$ in
order to avoid finite size effects. In Fig. 3(c) we can see the behavior
of the coefficient $\chi^{2}$ as a function of the multiplicity as calculated at 
$\tau_{ff}$ (full triangles) and $\tau_{a}$ (open triangles). The minimum
in $\chi^{2}$ signals the critical multiplicity: $m_{c}=59\pm 2$ for 
$\tau_{ff}$ and $m_{c}=76\pm 2$ for $\tau_{a}$. (The obtained values of $\tau$ 
from these fitting procedures will be reported in next sub-section).
These results agree very well with those obtained above via $NVM$ and $M_{2}$. 

The relation between the multiplicity and the energy of the system is plotted in 
Fig. 3(d). We can see that the critical multiplicities correspond mainly to
energies around $E=0.3\epsilon $, energy for which the collective motion
begins to be noticeable \cite{chernison}. It can be seen that there is a one to one
correspondence between the energy and the multiplicity and, moreover, starting
right bellow $m_{c}$ the relation between energy and multiplicity becomes
almost linear. This feature supports the use of multiplicity as a control
parameter. A similar behavior was also observed in percolation for the
relation between the multiplicity and the bond probability $p$.

All these signals show evidences of critical behavior not only at asymptotic
times, when fragments are experimentally accessible, but also at
fragmentation time, when the particles of the system are still interacting
and they are recognized by the ECRA algorithm.

\subsection{Critical exponents}

We now calculate four critical exponents: $\tau$, $\gamma$, $\sigma$, and $%
\beta$ at time of fragment formation ($\tau_{ff}$) and at asypmtotic time ($%
\tau_{a}$).

The calculation of $\tau$ is performed according to the procedure described
in the previous section, i.e., finding the fragment mass spectra that is
best fitted by a power law according to the scaling model in a given range of 
masses. From this calculations we got $\tau=2.18\pm 0.03$ for both times, 
$\tau_{ff}$ and $\tau_{a}$. The biggest fragment of each event was excluded 
when generating the average cluster distribution of a given multiplicity 
\cite{elliot}. We can see that the resulting value of $\tau $ does not depend 
on the fit range as long as we stay in the range $0.02A_{tot}<A<0.15A_{tot}$, 
where finite size effects are not important. The cluster mass distributions 
at the critical multiplicity and their corresponding fits are shown in Fig. 4(a) 
for $\tau_{ff}$ and Fig. 5(a) for $\tau_{a}$.

\vspace{0.8cm} 
\begin{figure}
\centerline{\epsfig{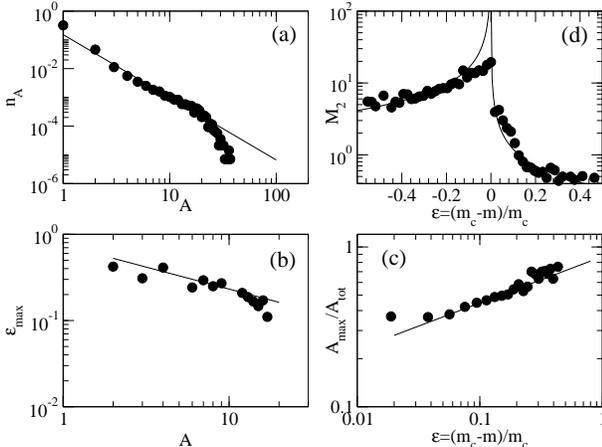}}
\caption{The critical exponents at $\tau_{ff}$: $\tau$ (a), $\sigma$ (b) ,
$\beta$ (c) and $\gamma$ (d). In (a) we plot the number of clusters per
particle, $n(A)$, as a function of the mass number $A$ and its
best fit (full line). In (b), the value of $\epsilon_{max}$ (defined in 
the text), is plotted against the mass number $A$ and its best fit (full line). 
The slope gives the value of $\sigma$. In (c) we plot the mass of the biggest 
fragment normalized to the size of the system ($A_{max}/A_{tot}$) as a 
function of $\epsilon=\frac{m_c-m}{m_c}$ and its fit giving the value of $\beta$.
In (d) we plot the $M_2$ as a function of $\epsilon$ and the 
curve with the value of $\gamma$ obtained in the fitting procedure (full line).} 
\end{figure}

We have seen in the previous section that the argument of the scaling
function, $z$, can be expressed in terms of $\sigma$ via equation (\ref
{eq:sigma}). Taking into account that $f(z)$ has a single maximum, we can
calculate $\sigma$ in terms of the multiplicity by looking for the value of 
$\epsilon$ ($\epsilon_{max}$) that maximizes the production of clusters of a given 
size $A$: 

\begin{equation}
n_A^{max}(\epsilon_{max})=q_0A^{-\tau}f(z_{max}),
\end{equation}
where the argument of $f$ is $z_{max}=A^{\sigma}\epsilon_{max}$. This can be
rewritten as: 
\begin{equation}
\epsilon_{max}=z_{max}A^{-\sigma}
\end{equation}

Therefore, by plotting $\epsilon_{max}$, as a function of $A$ we get $\sigma$ and 
$z_{max}$ \cite{elliot,stanley}. The results obtained are $\sigma=0.51\pm0.15$ for
$\tau_{ff}$ and $\sigma=0.64\pm0.18$ for $\tau_a$, and they are plotted in 
figures 4(b) and 5(b) respectively. Although the values of $\sigma$ at time of 
fragment formation and at fragmentation time do not agree as well as 
$\tau$, they do agree when error bars are considered.

We calculate then the $\beta$ exponent related to behavior of the order 
parameter, which in our case is the biggest fragment mass ($A_{max}$). The 
behavior in the limit $\epsilon \rightarrow 0$ in an infinite system is given by:

\smallskip 
\begin{equation}
(A_{\max }/A_{tot})\propto \epsilon^{\beta},\; \epsilon <0,  \label{eq:beta}
\end{equation}

where $A_{max}$ is the average mass of the biggest fragment at a given
multiplicity, $A_{tot}$ is the mass of the system and $\epsilon$ the distance to 
the critical point as defined above. In finite systems, \ref{eq:beta} is valid 
in an intermediate region of $\epsilon$ where finite size effects are negligible. 
The relations $(A_{max}/A_{tot})\hspace{0.2cm}vs\hspace{0.2cm}\epsilon$ are
displayed in figures 4(c) for $\tau_{ff}$ and 5(c) for $\tau_a$. The obtained 
values of $\beta$ are $0.29\pm 0.08$  and $0.28\pm 0.13$ respectively. 
These values are very close to $0.31$, which is the one corresponding 
to liquid-gas transition (3D-Ising Universality class ), and far from the
percolation one (0.45). However, we believe that the smallness of the system is 
responsible of the large errors in $\beta$ and of a possible sub-valuation of the 
exponent, as was noticed in \cite{elliot1}.

\vspace{0.7cm}
\begin{figure}
\centerline{\epsfig{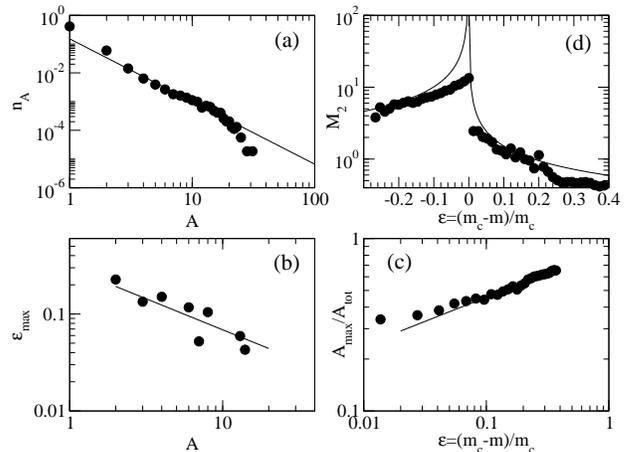}}
\caption{The same plot that in figure (4) but for asymptotic time $\tau_a$}.
\end{figure}

As we have shown in equation (\ref{eq:gamma}), the behavior of $M_{2}$ near the 
critical point can be described in terms of the critical exponent $\gamma$.
However the divergence predicted in this equation for the limit 
$\epsilon \rightarrow 0$, is valid for an infinite system. For a finite system, 
we have to find an intermediate region of $\epsilon $ where 
$M_{2}\propto \epsilon^{\gamma}$ (In this calculation, the biggest fragment of 
each event is removed when the multiplicity is below $m_c$). 
The procedure used to determine $\gamma $ is the following \cite{elliot}: first, 
we take a value of $m_{c}$ from the range given by $NVM$, $M_{2}$ and the best 
fitted mass spectrum. Then we determine two ranges of fitting
boundaries, one for the gas region ($\epsilon <0$, [$\epsilon_{g}^{min}:
\epsilon_{g}^{max}]$) and the other for the liquid region $(\epsilon >0,
[\epsilon_{l}^{min}:\epsilon_{l}^{max}])$. In each one of these regions,
we fit the $M_2(m)$ to a power law, saving the exponents $\gamma_{g}$ and 
$\gamma_{l}$, and their corresponding $\chi_{g}^{2}$ and 
$\chi_{l}^{2}$ for the gas and the liquid regions respectively. For each
value of $m_{c}$ we make several fitting procedures in a wide range of 
$\epsilon$, choosing those exponents ($\gamma_{g}$ and $\gamma_{l}$) that 
fulfil two conditions: first, the values should match each other within the 
error bars given by the fitting routine, and second, the $\chi^2$ of the fits 
should lie in the lowest twenty percent of the distribution.   
The values that satisfy these criteria were histogrammed and the average value 
($<\gamma>$) and the variance ($\sigma_{\gamma}$) were obtained from the
corresponding distribution . At $\tau_{ff}$ we got $\gamma =0.77 \pm
0.25 $ and at $\tau_{a},$ $\gamma =0.72\pm 0.33$. The results obtained are
displayed in figures 4(d) and 5(d) for $\tau_{ff}$ and $\tau_{a}$
respectively. We can see that the value of $\gamma$ is smaller than the one 
expected for a liquid-gas phase transition, although far away from the 
percolation value (see table I). 

In Table 1 we can see the values of the critical exponents calculated 
at time of fragment formation ($\tau_{ff}$) and at asymptotic times ($\tau_a$).
Moreover, the values of these exponents for two different universality classes, 
3D-Ising and Percolation, are also presented. It is clear from this table that:

1 - The time evolution after fragmentation time does not alter significatively 
the values of the critical exponents, i.e. the exponents obtained from 
asymptotic fragments distributions properly characterize the fragmentation process.

2 - The exponents are closer to the liquid-gas universality class than to
the percolation one. However, despite the strong evidence arisen in the previous 
subsection (see Fig. 3), we can not be conclusive about the critical 
behavior of the drop: the error bars in $\sigma$ avoid us to discern if the 
transition could be cast in some of the universality clasess presented; the value of 
$\beta$ is closer to 3D-Ising but it could be sub-valuated due to the smallness 
of the system, and finally $\gamma$ shows a value close to 3D-Ising, but a little 
bit lower than the expected value for this universality class.

We think that two main features of the studied system should  be taken into 
account in order to analyze the obtained results. First its finite size, that 
smoothes any signal of the possible phase transition and, consequently, induce 
quite large error bars for the calculated exponents. Second, even a certain degree 
of equilibration can be achieved at $\tau_{ff}$, non-equilibrium effects could 
still be noticeable (see \cite{cherison}) and be responsible of the 
disagreement between the obtained $\gamma$ exponent and the expected value from the  
3D-Ising Universality class.

\begin{table}[htbp]
\begin{center}
\begin{minipage}[t]{3.5in}
\centering
\caption{Critical exponents calculated at fragmentation and asymptotic 
time, and the corresponding to the 3DIsing (liquid-gas) and Percolation 
universality class [20,21]}.
\begin{tabular}{|l|l|l|l|l|}
 \hline
 Exponents  &  $\tau_{ff}$  &  $\tau_{a}$  & 3D-Ising & Percolation \\ \hline
 $\tau$  &  $2.18\pm 0.03$  &  $2.18\pm 0.03$  &  2.21  &  2.18   \\ \hline
 $\sigma$  &  $0.51\pm 0.15$  &  $0.64\pm 0.18$  &  0.64  &  0.45  \\  \hline
 $\beta$  &  $0.29\pm 0.08$  &  $0.28\pm 0.13$  &  0.33  &  0.41   \\   \hline
 $\gamma$  &  $0.77\pm 0.25$  &  $0.72\pm 0.33$  &  1.23  &  1.82   \\ \hline
\end{tabular}
\end{minipage}
\end{center}
\end{table}

\section{Conclusions}

In conclusion, we have calculated different signals of critical behavior at
fragmentation time as well as at asymptotic time. We have found that the 
dynamical evolution that follows the fragmentation process does not change the 
system criticallity features, i.e., the critical exponents values calculated at 
time of fragment formation. In this way, this work shows that it is possible 
to get the critical exponents that describes the physics involved in the 
fragmentation process from the mass distribution obtained experimentally.
On the other hand, we can not assure than the
process under study can be cast into the 3D-Ising universality class nor
into the percolation one. In particular, the $\gamma$ exponent is far too
low from the corresponding value for 3D-Ising class. This might suggest that
non-equilibrium effects play a crucial role at time of fragment formation.
Actually, at this time the system is in expansion and the collective motion is  
noticeable for excitation energies near the critical point, as was observed 
in the caloric curve of Ref.\cite{chernison}. 

\smallskip
We acknowledge partial financial support from UBA grant
EX139. P.B and A.C are fellows of the CONICET. C.O.D. is a member of the
Carrera del Investigador (CONICET)


\begin{references}
\bibitem{purdue}  A.S.Hirsch, A.Bujak, J.E.Finn, L.J.Gutay, R.W.Minich, N.T.Porile, 
R.P.Scharenberg, B.C. Stringfellow and F.Turkot, {\it Phys.Rev.C} {\bf 29}, 508
(1984).

\bibitem{gilkes}  M. L. Gilkes et al., {\it Phis. Rev.Lett. } {\bf 73}, 271 (1994).

\bibitem{latobona}  M. Belkacem, V. Latora and A. Bonasera, {\it Phys. Rev. C} 
{\bf 52}, 271 (1995).

\bibitem{nuovochi}  A. Bonasera, M. Bruno, C. O. Dorso and  P. F. Mastinu, {\it La
Rivista del Nuovo Cimento} {\bf 23}, 2 (2000)

\bibitem{dorlo}  J. A. Lopez and C. O. Dorso, {\it''Phase Transformation in Nuclear
Matter''}, World Scientific, (2000).

\bibitem{philip}  F. Gulminelli and P. Chomaz, {\it Phys. Rev. Lett.} {\bf 82}, 1402
(1999); J.M. Carmona, N. Michel, J. Richert, and P.Wagner, {\it Phys. Rev. C},{\bf 61} 037304 (2000).

\bibitem{elliot2}  J.B. Elliott, M.L.Gilkes, J.A.Hauger, A.S.Hirsch, E.Hjort, 
N.T.Porile, R.P.Scharenberg, B.K.Srivastava, M.L.Tincknell and P.G.Warren ,
{\it Phys. Rev. C} {\bf 55},1319 (1997).

\bibitem{nato}  J. B. Natowitz et al., {\it Phys. Rev. C} {\bf 65} (2002) 
034618.

\bibitem{dagostino}  M. D'Agostino et.al. {\it Nucl. Phys. A} {\bf 650}, 329 (1999).

\bibitem{elliot}  J. B. Elliott et al., 
{\it Phys. Rev. C} {\bf 62} (2000) 064603.

\bibitem{bauer}  M. Kleine Berkenbusch, W. Bauer et al.,
{\it Phys.Rev.Lett.} {\bf 88} (2002) 022701.

\bibitem{ecra93}  C. O. Dorso and J. Randrup, {\it Phys. Lett. B} {\bf 301}, 328 
(1993).

\bibitem{strador}  A. Strachan and C. O. Dorso, {\it Phys. Rev. C} {\bf 56},995
(1997).

\bibitem{strador2}  A. Strachan and C. O. Dorso, {\it Phys. Rev. C} {\bf 59},285
(1999).

\bibitem{chernodor}  A. Chernomoretz, C. O. Dorso and J. A. Lopez, {\it Phys. Rev. C}  
{\bf 64},044605 (2001).

\bibitem{aichelin}  C. O. Dorso and J. Aichelin, {\it Phys. Lett. B} {\bf 345},
197 (1995).

\bibitem{chernison}  A. Chernomoretz, M. Ison, S. Ortiz and C. O. Dorso,
{\it Phys. Rev. C} {\bf 64},024606 (2001).

\bibitem{gulmi}  F. Gulminelli, Ph. Chomaz, M. Bruno and M. D'Agostino,
{\it Phys.Rev. C} {\bf 65} (2002) 051601.

\bibitem{stauffer}  D. Stauffer and A. Aharoni, {\it ``Introduction to Percolation
Theory''}, 2nd edition, (Taylor and Francis, London, 1992).


\bibitem{stanley}  H. Nakanishi and H. E. Stanley, {\it Phys. Rev. B} {\bf 22}, 
2466 (1980).

\bibitem{dorbona}  C. O. Dorso, V. C. Latora and A. Bonasera , {\it Phys. Rev. C}, 
{\bf 60}, 034606 (1999).

\bibitem{elliot1}  J.B. Elliott, M.L.Gilkes, J.A.Hauger, A.S.Hirsch, E.Hjort, 
N.T.Porile, R.P.Scharenberg, B.K.Srivastava, M.L.Tincknell and P.G.Warren ,
\underline{PRC} {\bf 49},3185 (1994).

\bibitem{cherbado} A. Chernomoretz, P.Balenzuela and C.O.Dorso, preprint 
{\it arXiv:nucl-th/0203050}.

\end{references}
\end{document}